\documentclass[cameraready]{Interspeech}

\title{Dual-Branch Gated Fusion for Open-Set Audio Deepfake Source Tracing}

\author[affiliation={1,2}, orcid=0000-0003-2497-7687]{Awais}{Khan}
\author[affiliation={1}, orcid=0000-0003-4356-682X]{Kutub}{Uddin}
\author[affiliation={1,2}, orcid=0000-0002-7927-3436, correspondingauthor]{Khalid}{Malik}

\address{
    $^1$ College of Innovation and Technology, University of Michigan, Flint, MI, USA \\
    $^2$ ProbeTruth Inc., MI, USA 
}

\email{mawais@umich.edu, kutub@umich.edu, malik@probetruth.ai}
\usepackage{comment}
\usepackage{tikz}
\usetikzlibrary{positioning,arrows.meta,calc}
\usepackage{stfloats}
\usepackage{tabularx}
\usepackage{amsmath}
\usepackage{amssymb}
\usepackage{xcolor}
\usepackage{cite}
\hypersetup{hidelinks}

\begin{document}

\maketitle
\begin{abstract}
Attributing a synthetic utterance to its originating system remains an open challenge: closed-set models fail to reject unseen synthesizers and produce overconfident predictions. To address this, we propose a dual-branch gated fusion framework that pairs XLSR-53 with CORES, a 66-dimensional descriptor that, unlike prior Linear Filter Bank-only work, spans cepstral, oscillatory, rhythmic, energy, and spectral dimensions to capture complementary synthesis artifacts. Our analysis shows XLSR-53 remains discriminative in-domain (ID) while CORES generalizes stably under distribution shift (OOD), yet their naive concatenation fails due to SSL representational imbalance. To resolve this, an input-conditioned gate adaptively weights each branch under joint training with cross-entropy, an energy margin loss for ID/OOD separation, and a gate diversity term. On the MLAAD benchmark, our system achieves 97.6\% ID accuracy, 4.9\% EERc, and an 83.5\% relative reduction in FPR95 over the 2025 baseline.
\end{abstract}
 \vspace{-5mm}
\section{Introduction}
\label{sec:intro}
\textbf{Hearing is no longer believing.} This warning, once reserved for gossip and hearsay, has taken on an unsettling new dimension in the age of generative AI. In January 2024, a phone call impersonating the voice of US President Biden discouraged thousands of New Hampshire voters from attending the primary election, triggering a federal investigation~\cite{bbc_fake_biden_robocall_2024}. The threat has not abated: in January 2026, a businessman in the Swiss canton of Schwyz was deceived over a series of calls in which fraudsters cloned the voice of a trusted business partner, ultimately transferring several million Swiss francs before the deception was uncovered~\cite{deepfake_swiss_2026}. These incidents share a common thread that the attacker's identity remained hidden, not because the audio was undetectable, but because the tools to trace its origin were not in place. Knowing that a recording is synthetic is, increasingly, not enough. The more consequential question in digital forensics and accountability is which system produced it.

Audio deepfake detection~\cite{zhang2025audio,khan2023battling, uddin2025advbench} has reached a level of maturity where systems report equal error rates (EER) below 0.5\% on established benchmarks~\cite{tak2022automatic,li2025we,farooq2025transferable,khan2024frame,xie2026interpretable}. Source tracing, in contrast, which includes attributing a synthetic utterance to the specific TTS or voice conversion pipeline that generated it, is a harder and considerably less studied problem. Early work~\cite{zeng2023deepfake,xuan2025multilingual,koutsianos2025synthetic,xie2024generalized,wang2023npu,tian2023deepfake} framed it as multi-class classification over a fixed vocabulary of known systems. Granularity levels range from the model level~\cite{phukan2025investigating,lu2023detecting} to higher-level architecture components such as acoustic models and vocoders~\cite{zhang2024distinguishing,klein2024source, doan2025vib}. In particular, Klein et al.~\cite{klein2024source} demonstrated that jointly classifying the acoustic model and the vocoder attributes across the generation pipeline is a viable strategy, while Doan et al.~\cite{doan2025vib} introduced Variational Information Bottleneck regularization to improve generalization within this closed-set regime. However, both lines of work~\cite{klein2024source,doan2025vib} rest on an assumption that fails immediately in practice, that every synthesis system encountered at the deployment time was also seen during training.

The ecosystem of generative speech models grows faster than any fixed training set. The ADD2023~\cite{yi2023add} challenge Track 3 included only a single held-out out-of-distribution (OOD) system, and subsequent studies have reached at most five unseen classes~\cite{muller2022attacker,xie2025neural}. Interspeech 2025 responded by organizing a dedicated source tracing special session around the MLAAD protocol~\cite{MLAAD}, which positions 43 entirely unseen synthesis systems in the evaluation set, making it the most demanding open-set benchmark the field has seen to date. Underlying this challenge is a broader requirement for trustworthy AI: a source tracing system must reliably recognize what it does not know, and its confidence should naturally decay when confronted with synthesis pipelines outside its training distribution (OOD data). On this benchmark, the dominant approach is to fine-tune large self-supervised learning (SSL) representations such as wav2vec2.0 end-to-end with a discriminative backend such as AASIST (Interspeech special session baseline\footnote{\url{https://github.com/piotrkawa/audio-deepfake-source-tracing}}\label{fn:github}). Following this, Kulkarni et al.~\cite{kulkarni2025unveiling} show that XLSR-Conformer variants with deep metric learning exceed 95\% in-domain (ID) accuracy, while Klein et al.~\cite{klein2025openset} demonstrate that pairing a ResNet34 with large margin cosine loss and softmax energy-guided training achieves 8.3\% FPR95. Yet a fundamental challenge persists across these systems: the same contextual richness that makes SSL representations discriminative for known sources causes them to overcommit to the training distribution, assigning high confidence scores to utterances from synthesis pipelines the model has never seen. The best OOD-robust configuration reported in~\cite{kulkarni2025unveiling} is XLSR-53 with HYDRA (S5), which achieves its OOD advantage at the cost of 23 percentage points (pp) in in-domain accuracy, an unacceptable trade-off for operational deployment.

This conflict between ID classification and OOD detection is not a tuning problem; it reflects a deeper mismatch between what deep self-supervised learning (SSL) features are optimized for and what open-set rejection requires. Handcrafted spectral and temporal features encode lower-level signal properties that are more conservative under distribution shift: cross-lingual source tracing experiments~\cite{xuan2025multilingual} have shown that LFCC-based representations generalize more stably to unseen conditions precisely because they do not over-commit to high-level contextual abstractions~\cite{kulkarni2025unveiling}. The complementarity between deep SSL representations and handcrafted acoustic descriptors is well documented across speech tasks~\cite{schneider2019wav2vec}, yet it has not been deliberately exploited for OOD detection in source tracing. To test this directly, we first experiment with a fixed concatenation of the two feature streams. We observe, however, that joining a 1024-dimensional XLSR embedding with a 66-dimensional handcrafted vector without any conditioning causes the fused representation to be numerically dominated by the SSL component: the model overfits to the synthesis artifacts present in the training distribution and the handcrafted branch contributes negligibly, leaving the system with the same overconfidence on unseen systems that a single-stream SSL model exhibits~\cite{kulkarni2025unveiling}. This finding motivates an input-conditioned gating mechanism that dynamically reweights the two branches rather than their fusion as a fixed operation.

To address this, we propose a dual-branch gated fusion framework for open-set audio deepfake source tracing. One branch operates on frozen XLSR-53 embeddings that capture phonetic, prosodic, and speaker-level structure from large-scale multilingual pretraining; the other operates on a carefully designed 66-dimensional handcrafted descriptor we term CORES (Cepstral, Oscillatory, Rhythmic, Energy, and Spectral). Unlike prior handcrafted representations used in source tracing, which have typically relied on linear filter-bank (LFB) alone~\cite{klein2025openset}, CORES is purpose-designed to span the full range of signal-level artifacts that distinguish synthesized audio from natural speech and differentiate across synthesis pipelines. To the best of our knowledge, this five-dimensional combination has not been previously proposed for source tracing or deepfake detection, and it provides distribution-agnostic coverage of synthesis artifacts precisely because it encodes signal dimensions that deep SSL representations do not prioritize. The relative contribution of each branch is determined by a lightweight gating network conditioned on the input itself, trained jointly with an energy margin loss that drives ID/OOD score separation and a gate diversity term that prevents collapse toward a single-branch solution. At inference, this input-conditioned routing maintains strong closed-set classification without sacrificing open-set sensitivity, a balance that fixed-weight fusion and single-stream SSL architectures cannot simultaneously achieve.
 \vspace{-3mm}
\begin{figure*}[!b]
\vspace{-15pt}
\centering
\includegraphics[width=\textwidth]{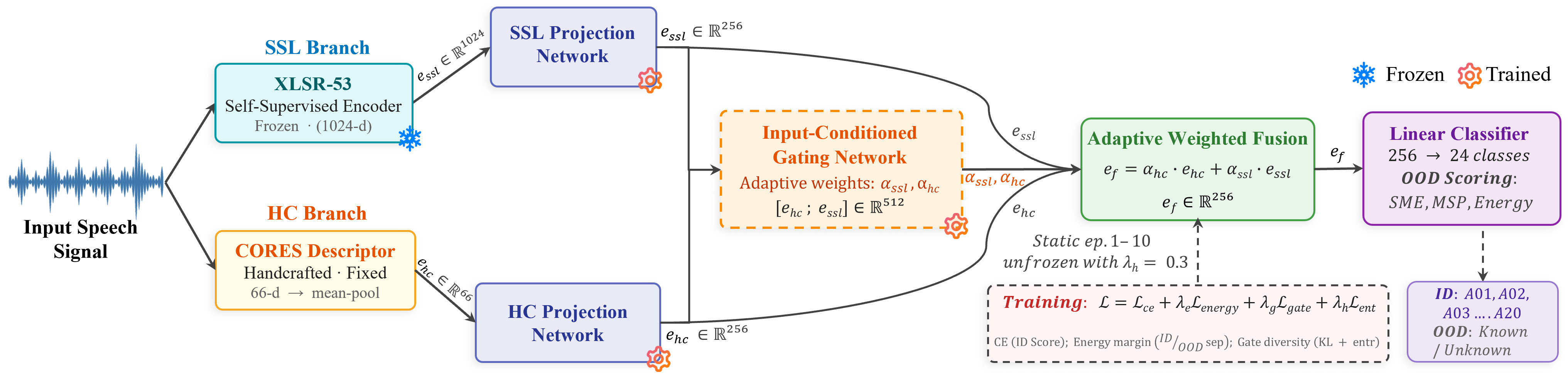}
\caption{Proposed dual-branch gated fusion framework for open-set audio deepfake source tracing. The gating network adaptively balances contextual SSL representations and low-level CORES to improve attribution robustness under unseen generative systems.}
\label{fig:flow}
\end{figure*}
\section{Proposed Method}
\label{sec:proposed}
\vspace{-4pt}
Our approach is based on a central observation: synthesis artifacts span multiple representational levels, affecting fine spectral cues, harmonic and temporal dynamics, and higher-level phonetic structure. No single feature family captures these reliably across both seen and unseen systems. We therefore combine two complementary front-ends: a handcrafted acoustic descriptor and a frozen self-supervised representation, and replace fixed fusion with an input-conditioned gating module that adaptively weights each branch based on the sample’s distributional properties. The overall framework is illustrated in Figure~\ref{fig:flow}.
\vspace{-3mm}
\subsection{Feature Extraction}\vspace{-3pt}
We extract two representations from each utterance. The first is produced by a frozen XLSR-53 encoder~\cite{babu2021xls}, pretrained on 56{,}000 hours of multilingual speech. We freeze the encoder to preserve generalized representations and prevent overfitting on the training subset, then mean-pool the final hidden states to yield $\mathbf{x}_{\mathrm{ssl}}\in\mathbb{R}^{1024}$. The second is a 66-dimensional handcrafted descriptor $\mathbf{x}_{\mathrm{hc}}\in\mathbb{R}^{66}$, referred to as CORES, which captures complementary signal-level synthesis artifacts. It comprises five complementary dimensions: \textit{Cepstral:} 13 MFCCs with first ($\Delta$) and second-order ($\Delta\Delta$) derivatives (39-d); \textit{Oscillatory:} chroma features capturing pitch-class tonal structure (14-d); \textit{Rhythmic:} zero-crossing rate reflecting temporal onset patterns (1-d); \textit{Energy:} RMS energy (1-d); and \textit{Spectral:} spectral centroid, bandwidth, and roll-off (3-d), spectral contrast (7-d), and spectral flatness (1-d). Frame-level values are 
mean-pooled to yield the 66-dimensional utterance descriptor. While handcrafted features have been employed in prior source tracing~\cite{klein2025openset} and deepfake detection~\cite{khan2023securing,todisco2017constant}, most commonly as LFCC or MFCC descriptors in isolation~\cite{xuan2025multilingual, ali2025audio,khan2023spotnet}, CORES deliberately combines five complementary signal dimensions to cover the breadth of artifacts introduced across different synthesis architectures. We do not claim novelty of individual components, but of their deliberate five-dimensional combination for distribution-agnostic coverage of synthesis artifacts. Unlike the SSL representation, it requires no learned parameters, making it invariant to training distribution and conservatively stable under distribution shift.
 \vspace{-3mm}
\subsection{Dual-Branch Expert Architecture}\vspace{-3pt}
\label{sec:2.2}
Within the proposed system, each feature type is processed by a dedicated projection network. Both map their inputs through two fully-connected layers (hidden dim 512, BatchNorm, ReLU, dropout $p{=}0.3$) into a shared 256-dimensional embedding space: $\mathbf{e}_{\mathrm{hc}}, \mathbf{e}_{\mathrm{ssl}} \in \mathbb{R}^{256}$. Direct concatenation is insufficient, as the SSL input is $15{\times}$ larger than CORES before projection and retains substantially higher gradient energy afterward, causing the model to route almost entirely through the SSL branch. We validate this behavior via ablation in Section~\ref{sec:experiments}. \vspace{-6mm}
\subsection{Input-Conditioned Gating}\vspace{-3pt}
To resolve the dominance problem, we introduce a lightweight gating network that computes a soft branch allocation from the input itself.  Given the concatenated expert embeddings $[\mathbf{e}_{\mathrm{hc}};\, \mathbf{e}_{\mathrm{ssl}}] \in \mathbb{R}^{512}$, the gate applies a two-layer gating network ($512 \to 128 \to 2$, ReLU, dropout $p = 0.2$) followed by softmax:
\begin{equation}
[\mathbf{\alpha}_{\mathrm{hc}}, \mathbf{\alpha}_{\mathrm{ssl}} ]= \mathrm{softmax}[
\mathbf{W}_2\,\mathrm{ReLU}\left(\mathbf{W}_1\,(\mathbf{f}_{e_{\mathrm{hc}}}, \mathbf{f}_{e_{\mathrm{ssl}}})\right)]
\label{eq:gate}
\end{equation}
The fused embedding is the gated weighted sum:
\begin{equation}
  \mathbf{e}_{\mathrm{fused}}
  = \alpha_{\mathrm{hc}} \cdot \mathbf{e}_{\mathrm{hc}}
  + \alpha_{\mathrm{ssl}} \cdot \mathbf{e}_{\mathrm{ssl}} \\\
  \in \\\ \mathbb{R}^{256}
  \label{eq:fusion}
\end{equation}
A linear classifier maps $\mathbf{e}_{\mathrm{fused}}$ to logits over 24 ID classes.
 \vspace{-2mm}
\subsection{Training Objectives}\vspace{-2.5pt}
We combine three complementary losses that jointly encourage in-distribution classification, out-of-distribution separability, and input-dependent gate specialization. \\
\noindent\textbf{Source classification.}
We first apply cross-entropy loss with label smoothing ($\varepsilon = 0.15$) to the ID logits, which anchors the classifier to discriminate among the 24 seen synthesis systems.
\begin{equation}
\mathcal{L}_{\mathrm{CE}} = - \sum_{i: y_i \neq -1} \sum_{c=1}^{C} q_c^{(i)} \log \hat{p}_c^{(i)}, 
q_c^{(i)} = (1-\epsilon)\mathbf{1}[y_i = c] + \frac{\epsilon}{C},
\end{equation}
\noindent\textbf{Energy margin loss.}
$\mathcal{L}_{\mathrm{CE}}$ does not provide a signal for rejecting unseen systems. We therefore adopt the energy margin loss of Liu et al.~\cite{liu2020energy}, using Dev-split OOD samples as auxiliary data. Defining $E(\mathbf{x}){=}{-}\log\sum_k e^{z_k}$, we penalize ID samples with energy above $m_{\mathrm{in}}$ and OOD samples below $m_{\mathrm{out}}$:
\begin{equation}
\begin{aligned}
\mathcal{L}_{\mathrm{energy}}&=\mathbb{E}_{\mathbf{x}_{\mathrm{in}}}[\mathrm{ReLU}E(\mathbf{x}_{\mathrm{in}})-m_{\mathrm{in}})] \\ &\quad + \mathbb{E}_{\mathbf{x}_{\mathrm{out}}}
[\mathrm{ReLU}(m_{\mathrm{out}}-E(\mathbf{x}_{\mathrm{out}}))]
\end{aligned}
\label{eq:energy}
\end{equation}
with margins $m_{\mathrm{in}} = -15.0$ and $m_{\mathrm{out}} = -2.0$. The loss is inactive once the margins are satisfied, providing a stable training signal without continuously distorting the classifier.\\
\noindent\textbf{Gate diversity loss.}
Without explicit regularization, the gate collapses toward the SSL branch. To counter this, we maximize the KL divergence between batch-mean gate distributions over ID and OOD samples to force input-dependent routing:
\begin{equation}
  \mathcal{L}_{\mathrm{gate}}
  = -D_{\mathrm{KL}}(\bar{\boldsymbol{\alpha}}_{\mathrm{id}}
      \,|| 
      \bar{\boldsymbol{\alpha}}_{\mathrm{ood}})  
  \label{eq:gate_kl}
\end{equation}
\begin{equation}
D_{\mathrm{KL}}(p||q) = \sum_k p_k \log \frac{p_k}{q_k}
\end{equation}
A gate entropy term $\mathcal{L}_{\mathrm{ent}} = \sum_k \alpha_k \log \alpha_k$ prevents degenerate
one-hot routing. Combining all three terms, our total training objective is:
\begin{equation}
  \mathcal{L}
  = \mathcal{L}_{\mathrm{ce}}
  + \lambda_e\,\mathcal{L}_{\mathrm{energy}}
  + \lambda_g\,\mathcal{L}_{\mathrm{gate}}
  + \lambda_h\,\mathcal{L}_{\mathrm{ent}}
  \label{eq:total}
\end{equation}
with $\lambda_e = 0.5$, $\lambda_g = 0.05$, and $\lambda_h = 0.3$. \vspace{-2mm}
\subsection{Implementation Details}\vspace{-2pt}
All models are trained for 150 epochs using AdamW (lr $= 10^{-4}$, weight decay $= 10^{-4}$) with cosine annealing to a minimum of $5 \times 10^{-6}$, batch size 128, and gradient clipping at $\ell_2$ norm 5.0. Dev-split OOD samples serve as auxiliary out-of-distribution data for $\mathcal{L}_{\mathrm{energy}}$ and $\mathcal{L}_{\mathrm{gate}}$, requiring no external corpus. We find that activating $\mathcal{L}_{\mathrm{energy}}$ before the expert branches have formed stable representations causes the gate to collapse toward the SSL branch from the first epoch, preventing the handcrafted branch from contributing meaningfully. We address this with a \emph{gate freeze} strategy: the gating network parameters are frozen for the first 10 epochs, during which $\mathcal{L}_{\mathrm{ce}}$ trains both expert branches independently. The gate is then unfrozen with a strong entropy regularization weight ($\lambda_h = 0.3$) to enforce complementary routing before the gate settles. Checkpoints are saved independently for each scorer based on the lowest FPR95 on the Dev set.
\subsection{OOD Detection at Inference}\vspace{-5pt}
At inference, OOD rejection requires no retraining: we apply three post-hoc scoring functions directly to the classifier logits. Each scorer assigns a scalar confidence score; an utterance is flagged as OOD when its score falls below a threshold $\tau$ set on the Dev set, and attributed to one of the 24 ID classes otherwise. Energy scoring~\cite{liu2020energy} uses the log-sum-exp of raw logits, Softmax Energy (SME)~\cite{klein2025openset} applies softmax first to accentuate the logit skew of confidently classified ID samples, and Maximum Softmax Probability (MSP) uses the negative peak softmax probability. In all cases, ID samples produce more negative scores than OOD samples, reflecting higher classifier confidence. Results are reported in Table~1; we adopt SME as our primary scorer. \vspace{-2mm}
\begin{table}[t]
\centering
    \caption{OOD detection performance under three post-hoc scoring functions on the MLAAD Eval set (epoch~122 checkpoint, ID Acc = 97.65\%). AUROC: Area Under the ROC Curve. All scorers use $T = 1.0$.}
    \vspace{-2mm}
\label{tab:scorer_comparison}
\setlength{\tabcolsep}{3.0pt}
\renewcommand{\arraystretch}{1.05}
\begin{tabular}{lcccc}
\toprule
\textbf{Scorer}
  & \textbf{AUROC} $\uparrow$
  & \textbf{FPR95}\%$\downarrow$
  & \textbf{OOD-EER}\%$\downarrow$
  & \textbf{EERc}\%$\downarrow$ \\
\midrule
Energy~\cite{liu2020energy}  & 0.796 & 21.2  & 24.1  & 13.23         \\
SME~\cite{klein2025openset}    & \textbf{0.965} & \textbf{10.4} & \textbf{7.62} & \textbf{4.98} \\
MSP & 0.963 & 10.5  & 7.71  & 5.03         \\
\bottomrule
\end{tabular}
\vspace{-3mm}
\end{table}
\begin{table}[t]
  \caption{MLAAD source tracing protocol statistics.}
  \label{tab:mlaad}
   \vspace{-2mm}
     \raggedright
  \setlength{\tabcolsep}{4.5pt}
  \begin{tabular}{lrrrr}
    \toprule
    Set   & ID Arch. & ID Samples & OOD Arch. & OOD Samples \\
    \midrule
    Train & 24 & 11{,}000 & ---  & ---    \\
    Dev   &  8 &  4{,}800 &  17  & 7{,}200 \\
    Eval  & 21 & 13{,}591 &  43  & 20{,}309 \\
    \bottomrule
  \end{tabular}
  \vspace{-5mm}
\end{table}
\vspace{-5pt}
\section{Experiments and Results}\vspace{-1pt}
\label{sec:experiments}
\subsection{Experimental Setup}
We evaluate on the MLAAD\footnote{\url{https://deepfake-total.com/mlaad}.} source tracing protocol~\cite{MLAAD}, which comprises 83 TTS systems across 26 languages partitioned into training, development, and evaluation splits (Table~\ref{tab:mlaad}). The evaluation set contains 43 entirely unseen synthesizers across 26 languages. To enrich OOD coverage, we augment both training ID samples and the Dev-OOD auxiliary split $5\times$ using the baseline augmentation pipeline~\cite{kawa_audio_deepfake}, codec distortion, additive noise from MUSAN~\cite{musan2015}, and reverberation from RIRs\footnote{\url{https://www.openslr.org/28/}} NOISES, raising the effective OOD auxiliary pool to 36,000 utterances. Crucially, the original (non-augmented) Dev-OOD samples are withheld from training; only transformed copies are used, to prevent any label leakage from the development distribution. We report three metrics: ID accuracy (seen-system classification), FPR95~\cite{klein2025openset} (OOD false positive rate at 95\% ID true positive rate), and EERc~\cite{klein2025openset} (joint metric counting an ID sample correct only when both correctly classified and accepted as in-distribution; lower is better). 
 \vspace{-2.0mm}
\subsection{Comparative Results}\vspace{-3pt}
Table~\ref{tab:results_klein} compares our system against two direct baselines. The dominant open-set approach is Klein et al.~\cite{klein2025openset}'s ResNet34 with Large Margin Cosine Loss (LMCL) and SME-guided training (318M parameters), which achieves 8.3\% FPR95 with copy-synthesis augmentation. The second is the publicly available Wav2Vec2.0-AASIST baseline~\cite{kawa_audio_deepfake} released with the Interspeech 2025 special session and establishes the base result at 63.0\% FPR95. Our Dual-Branch Gated Fusion achieves 97.6\% ID accuracy, 4.9\% EERc, and 10.4\% FPR95 using only Dev-OOD auxiliary data and no copy-synthesis, an 83.5\% relative reduction in FPR95 over the reference system and a 2.8\% relative improvement over the Klein et al. no-augmentation baseline~\cite{klein2025openset}. This shows that the improvement stems from feature-level complementarity rather than model scale, achieved without fine-tuning any SSL weights. Moreover, we observe that the gate assigns $\bar{\alpha}_{\mathrm{ssl}}{=}0.617$ for ID samples and shifts to $\bar{\alpha}_{\mathrm{ssl}}{=}0.587$ for OOD samples, consistent with our hypothesis that handcrafted features provide more stable cues under distribution shift.\\
\begin{table}[t]
  \centering
  \caption{Results on MLAAD Eval set. OOD metrics use the SME score.
           Aug: codec~+~reverb augmentation. $^\dagger$: Best checkpoint, –: not reported, BL: Baseline Reference system, publicly available at the special session repository.}
            \vspace{-3mm}
  \label{tab:results_klein}
  \setlength{\tabcolsep}{3pt}
  \begin{tabular}{lcccc}
    \toprule
    \textbf{System} & \textbf{Aux / Aug}
      & \textbf{ID Acc\%}${\bm{\uparrow}}$
      & \textbf{EERc\%}${\bm{\downarrow}}$ 
      & \textbf{FPR95\%}${\bm{\downarrow}}$  \\
    \midrule
    \multicolumn{5}{l}{\textit{Klein et al.: ResNet34+LMCL (318M params)}} \\
    B1~\cite{klein2025openset}         & None / \texttimes          & 95.7 &  9.0 & 10.7 \\
    B2~\cite{klein2025openset}         & None / \checkmark          & 95.8 &  8.8 &  9.9 \\
    B3~\cite{klein2025openset}    & ASV-CS / \checkmark      & 95.5 & 8.1 & 8.3 \\
    \midrule
    BL~\cite{kawa_audio_deepfake} & None / \texttimes & 85.0 & -- & 63.0 \\
    \midrule
    \multicolumn{5}{l}{\textit{Ours: Dual-Branch Gated Fusion (XLSR-53+CORES)}} \\
    Ours$^\dagger$ & Dev-OOD / \texttimes & \textbf{97.6} & \textbf{4.9} & 10.4 \\
    \bottomrule
  \end{tabular}
  \vspace{-3mm}
\end{table}
The gate behavior at convergence explains the robustness under OOD conditions. For ID samples, routing favors XLSR-53 due to its stronger in-distribution discrimination. Under distribution shift, $\bar{\alpha}_{\text{ssl}}$ decreases modestly but consistently, with a corresponding rise in the handcrafted branch, providing artifact-sensitive coverage absent in SSL features outside the training domain. This asymmetric routing is not manually imposed but emerges from the gate diversity loss $\mathcal{L}_{gate}$, confirming complementary rather than redundant roles.\footnote{Adding ASVspoof5 as auxiliary OOD data degraded performance; excessive source diversity led to gate collapse toward a single branch, suppressing adaptive routing and reducing open-set gains.} 
\begin{table}[t]
  \centering
  \caption{Comparison with Kulkarni et al.~\cite{kulkarni2025unveiling} on MLAAD.
           OOD EER is standard EER (not EERc). $^\dagger$: Best checkpoint, $*$: augmentation.
           Params. in millions (M).}
            \vspace{-3mm}
  \label{tab:results_kulkarni}
  \setlength{\tabcolsep}{3pt}
  \begin{tabular}{lc|cc|cc}
    \toprule
    & & \multicolumn{2}{c|}{\textbf{In-Domain Eval}}
      & \multicolumn{2}{c}{\textbf{OOD Eval}} \\
    \textbf{System} & \textbf{Params}
      & \textbf{Acc\%}${\bm{\uparrow}}$ & \textbf{F1\%}${\bm{\uparrow}}$
      & \textbf{Acc\%}${\bm{\uparrow}}$ & \textbf{EER\%}${\bm{\downarrow}}$  \\
    \midrule
    Baseline~\cite{kawa_audio_deepfake}   & 317.8 & 83.4 & 83.3 & 26.5 & 73.5 \\
    S1.~\cite{kulkarni2025unveiling}  & 318.1 & 90.7 & 91.1 & 37.7 & 61.5 \\
    S2.~\cite{kulkarni2025unveiling}  & 318.1 & 95.1 & 95.2 & 39.5 & 59.2 \\
    S3.~\cite{kulkarni2025unveiling}        & 318.1 & 95.1 & 95.3 & 38.9 & 61.1 \\
    S4.~\cite{kulkarni2025unveiling}        & 318.1 & 79.9 & 77.0 & 39.2 & 60.1\\
    S5.~\cite{kulkarni2025unveiling}     & 319.7 & 72.0 & 69.8 & 44.8 & 55.1 \\
    E1: S2+S3~\cite{kulkarni2025unveiling}   & --    & 95.6 & \textbf{95.7} & 38.0 & 61.9 \\
    E2: S3+S4~\cite{kulkarni2025unveiling}   & --    & 94.2 & 94.2 & 32.6 & 69.2 \\
    E3: S3+S5~\cite{kulkarni2025unveiling}   & --    & 94.6 & 94.6 & 38.0 & 61.3 \\
    E4: S4+S5~\cite{kulkarni2025unveiling}   & --    & 80.5 & 77.8 & 39.6 & 60.3\\
    \midrule
    Ours$^\dagger$ & \textbf{ 0.897} & \textbf{97.6} & 91.2 & \textbf{94.3} & \textbf{7.6} \\
    \bottomrule
  \end{tabular}
   \vspace{-5mm}
\end{table}
Table~\ref{tab:results_kulkarni} provides a second axis of comparison with Kulkarni et al.~\cite{kulkarni2025unveiling}, enabling direct efficiency analysis across architectures despite the absence of open-set FPR95/EERc metrics. Our model achieves 97.6\% ID accuracy and 91.2\% macro F1 with 0.9M parameters, outperforming the 317.8M W2V2-AASIST baseline~\cite{kawa_audio_deepfake} by 14.2 pp in accuracy and 7.9 pp in F1, while approaching the best 318M XLSR-Conformer variants at over $350\times$ fewer parameters.

On OOD Eval, we obtain 94.3\% OOD accuracy and 7.6\% EER, surpassing all systems, including the OOD-focused XLSR-HYDRA (S5), by 49.5 pp in OOD accuracy and 47.5 pp in EER, while also exceeding S5 in ID accuracy by 25.6 pp, resolving the ID/OOD trade-off inherent to single-stream models. This improvement stems from adaptive gating that routes OOD samples toward the artifact-sensitive handcrafted branch, combined with energy margin training that explicitly separates ID and OOD score distributions. Unlike prior systems, which treat OOD behavior as incidental, our open-set performance is a designed property achieved without sacrificing ID accuracy.
\vspace{-4pt}
\subsection{Ablation: Why Gating Is Necessary}
\begin{figure}[t]
  \centering
  \includegraphics[width=\columnwidth]{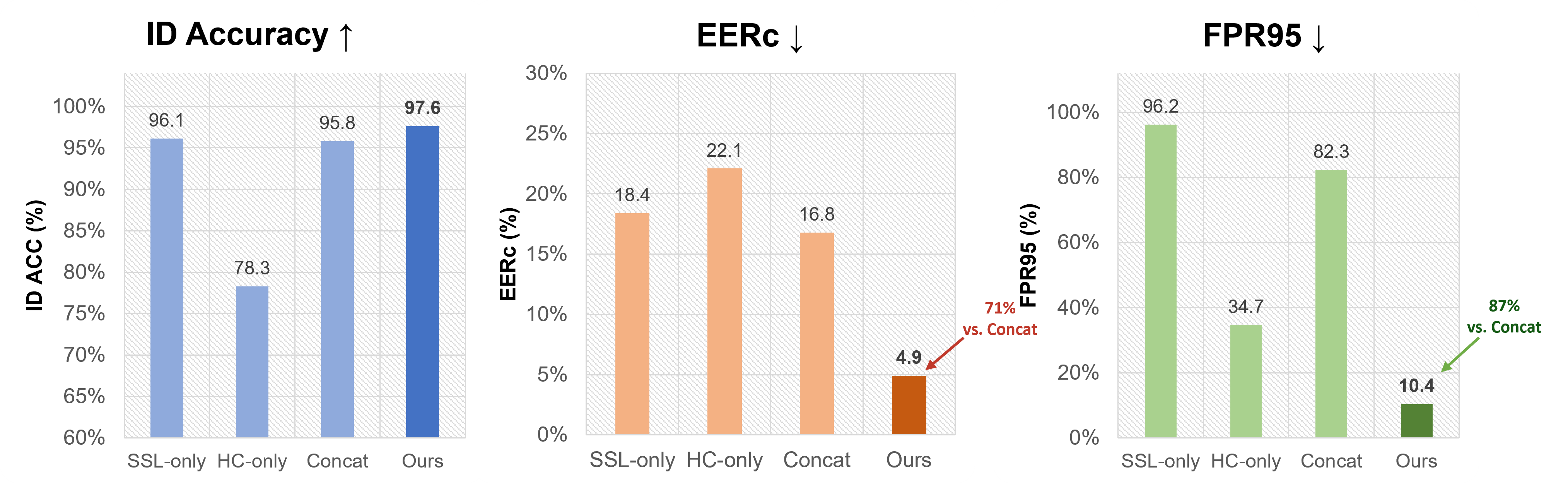}
  \vspace{-18pt}
  \caption{Ablation on MLAAD Eval. Single-branch and naive fusion baselines all fail OOD
           rejection (FPR95 $\geq$ 34\%). Adaptive gating resolves the ID/OOD conflict,
           achieving 87\% relative FPR95 vs. naive concatenation at no ID accuracy cost.}
  \label{fig:ablation}
   \vspace{-6mm}
\end{figure}

\vspace{-2pt}
Figure~\ref{fig:ablation} isolates the contribution of adaptive fusion. An AASIST backend trained on SSL features alone achieves competitive ID accuracy (96.1\%) but collapses at OOD rejection: FPR95 $=$ 96.2\%, confirming that the SSL energy landscape is flat across the seen/unseen boundary. HC-only training reduces FPR95 to 34.7\%, a 62 pp gain, but ID accuracy falls to 78.3\%, an unacceptable loss of classification power. Naively concatenating both branches with AASIST recovers ID accuracy (95.8\%) but leaves FPR95 at 82.3\%: the 1024-d SSL embedding numerically dominates the 66-d handcrafted vector, and the fused representation inherits the same overconfidence as SSL-only. Only the gated fusion with explicit energy margin training and gate diversity regularization simultaneously achieves strong ID accuracy and OOD rejection, reducing FPR95 by 87\% and EERc by 71\% relative to naive concatenation. This confirms the dominance effect described in Section~\ref{sec:2.2}: without input-conditioned weighting, gradient energy from the SSL branch overwhelms CORES regardless of projection dimensionality. \vspace{-4pt} 
\vspace{-3pt}
\section{Discussion and Future Work}
\vspace{-3pt}
Our results show that exploiting feature complementarity, rather than model scale, enables competitive open-set source tracing. Ablations confirm that neither SSL nor CORES alone, nor naive concatenation, resolves the ID/OOD trade-off; adaptive gating with energy margin training is essential. The consistent gate shift toward CORES under distribution shift further indicates that routing captures a genuine distributional property, offering a lightweight diagnostic for OOD detection without retraining. A key negative finding is that adding ASVspoof5 auxiliary data degraded performance due to gate collapse under high source diversity, indicating finite regularization capacity of the diversity loss. Curriculum-based auxiliary selection that gradually increases OOD diversity is a natural next step. Future work includes parameter-efficient adaptation of XLSR-53 via low-rank adaptation (LoRA) to reduce the FPR95 gap without sacrificing the efficiency that distinguishes our approach. Second, extending the framework to multi-label attribution, jointly predicting the acoustic model and the vocoder, would increase forensic granularity while retaining the open-set rejection capability.
\section{Acknowledgment}
This material is based upon work supported by the National Science Foundation (NSF) under Grant No.~2409577. Any opinions, findings, and conclusions or recommendations expressed in this material are those of the authors and do not necessarily reflect the views of the NSF.
\section{Use of Generative AI Disclosure}
Generative AI tools were used exclusively for grammar correction and linguistic refinement of the manuscript. No AI assistance was used in the conception, design, execution, or analysis of the research presented in this paper.
\bibliographystyle{IEEEtran}
\bibliography{mybib}
\end{document}